\documentclass[a4paper,amsfonts, amssymb, amsmath, reprint, showkeys, nofootinbib, superscriptaddress]{revtex4-2}
\usepackage[english]{babel}
\usepackage[utf8]{inputenc}
\usepackage{amsthm}
\usepackage{mathtools}
\usepackage{physics}

\usepackage{graphicx}
\usepackage{adjustbox}
\usepackage{placeins}
\usepackage[T1]{fontenc}
\usepackage{lipsum}
\usepackage{csquotes}
\usepackage[colorinlistoftodos, color=green!40, prependcaption]{todonotes}

\usepackage{xcolor}

\usepackage{hyperref}
\hypersetup{
    colorlinks=true,       % false: boxed links; true: colored links
    linkcolor=blue,        % color of internal links
    citecolor=blue,        % color of links to bibliography
    filecolor=blue,        % color of file links
    urlcolor=blue,         % color of external links
    runcolor=blue
}

\begin{document}

\title{Laser-assisted binding of ultracold polar molecules with Rydberg atoms \\in the van der Waals regime }

\author{Vanessa Olaya}
\affiliation{Department of Physics, Universidad de Santiago de Chile, Av. Ecuador 3493, Santiago, Chile}

\author{Jes\'{u}s P\'{e}rez-R\'{i}os}
\affiliation{Fritz-Haber-Institut der Max-Planck-Gesellschaft, Faradayweg 4-6, 14195 Berlin, Germany}

\author{Felipe Herrera}
\email{felipe.herrera.u@usach.cl}
\affiliation{Department of Physics, Universidad de Santiago de Chile, Av. Ecuador 3493, Santiago, Chile}
\affiliation{ANID-Millennium Institute for Research in Optics, Chile}

\begin{abstract}
We study ultracold long-range collisions of heteronuclear alkali-metal dimers with a reservoir gas of alkali-metal Rydberg atoms in a two-photon laser excitation scheme. In a low density regime where molecules remain outside the Rydberg orbits of the reservoir atoms, we show that the two-photon photoassociation (PA) of the atom-molecule pair into a long-range bound trimer state is efficient over a broad range of atomic Rydberg channels.  As a case study, we obtain the PA lineshapes for the formation of trimers composed of KRb molecules in the rovibrational ground state and excited Rb atoms in the asymptotic Rydberg levels $n^{2}S_j$ and $n^{2}D_j$, for $n=20-80$. We predict atom-molecule binding energies in the range $10-10^3$ kHz for the first vibrational state below threshold. The average trimer formation rate is order $10^8\, {\rm s}^{-1}$  at 1.0 $\mu$K, and depends weakly on the principal quantum number $n$. Our results set the foundations for a broader understanding of exotic long range collisions of dilute molecules in ultracold atomic Rydberg reservoirs.
\end{abstract}


\maketitle

\section{Introduction}

Rydberg atoms have a pivotal role in several areas of atomic, molecular, and optical physics \cite{Gallagher2005rydberg}, such as the implementation of novel quantum information protocols~\cite{Saffman2010}, quantum simulation of many-body Hamiltonians~\cite{Many_body_Rydberg}, the study of impurity physics, non-linear quantum optics, and ultracold chemistry~\cite{Michael2016b,JPRBook}. Many of these applications rely on controlling the exaggerated dipolar properties of Rydberg states in order to tailor the resulting interatomic forces. Rydberg excitations are generally distinguished via their characteristic lineshape, which in the case of an ultracold atomic gas depends on the Rydberg electron-background atom interaction~\cite{Liebisch2016}. At high densities, the Rydberg lineshape is affected by  collisional processes such as the formation of ultralong-range Rydberg molecules~\cite{Greene2000,Bendkowsky2009,Hamilton2002,Khuskivadze2002,Gaj2014,Trilobite,Butterlfy,ULR_1,ULR_2,ULR_3,Eiles2016}, and possible polaron effects when the Rydberg excitation behaves as an impurity in a dense atomic reservoir~\cite{Rydberg_Polaron,Rydberg_Polaron2}.

Recent studies have focused on the interaction of one or more diatomic molecules with a Rydberg alkali-metal atom in dense samples where molecules are likely to be inside the orbit of the Rydberg electron~\cite{Ferez2017,Ferez2020}. In addition to fundamental studies on their collisional properties, these molecule-Rydberg atom systems have been proposed as novel tools for quantum state preparation~\cite{Kuznetsova2011,Kuznetsova2016}. The formation of Rydberg {\it bimolecules} with binding energies in the GHz regime and kilo-Debye permanent dipole moments, has also been predicted~\cite{Ferez2021}. Bimolecules are formed by the interaction of a low-energy diatomic molecule with the outer electron of a Rydberg molecule, also at high gas densities.

\begin{figure*}[ht]
\centering
\includegraphics[width=0.9\textwidth]{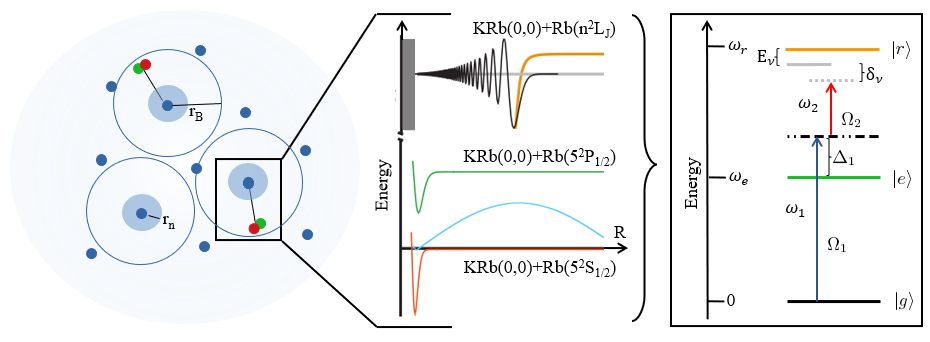}  
\caption{{\bf Photoassociation of a Rydberg atom and a diatomic molecule into an ultra-long range Rydberg trimer}. Left: Scheme of the atomic and molecular diluted gases, where the molecule can be found within the Rydberg blockade radius $r_B$, but not inside the Rydberg orbit radius $r_n$. Center: Interacting channels and their potentials in the PA process, showing the relevant scattering and bound state wavefunctions. The process efficiency depends on the overlap between the initial scattering state [KRb(0,0)+Rb($5^2S_{1/2}$)] and the final near-threshold vibrational state of the electronically excited trimer [KRb(0,0)+Rb($n^2L_j$)]. The substantial difference of van der Waals lengths between the initial and final states is highlighted. Right: Two-photon scheme for exciting Rb atoms into a target Rydberg state, where atomic states $|g\rangle, |e\rangle, |r\rangle$ are connected through the lasers $\omega_1$ and $\omega_2$, giving the two-photon detuning $\delta_\nu=\omega_1+\omega_2-(\omega_r-E_\nu))$ from the trimer bound level $E_\nu$. Large intermediate detunings $\Delta_1$ from the excited level $5^2P_{1/2}$ are needed to prevent gas heating via light scattering.}
\label{Fig_stellar}.
\end{figure*}

In this work, we study the two-photon photoassociation spectrum of heteronuclear alkali-metal dimers with alkali metal Rydberg atoms at ultracold temperatures. We focus on the regime where the molecular density is much smaller than the density of Rydberg atoms. We estimate the latter by requiring that at most one atom resides within the Rydberg blockade radius $r_B$, which depends on the van der Waals interaction between atoms in identical Rydberg levels~\cite{Saffman2010}.  Given a laser-dependent Rydberg excitation fraction $f<1$, the critical ground  atom density $\rho_B$ below which we can ignore blockade effects can be estimated  from $f\times \rho_B\times (4/3)\pi r_B^3=1$. Assuming weak dressing $f\sim 10^{-2}$ and $r_B\sim 1\,\mu{\rm m}$ \cite{Singer_2005}, the relevant atom densities for this work $\rho_B\sim 10^{10}\,{\rm cm}^{-3}$  are thus compatible with magneto-optical trapping~\cite{SingerK,Tong}.  

As we illustrate in Fig. \ref{Fig_stellar}, the density of alkali-metal dimers in the reservoir of ground state and Rydberg state atoms is low enough to make the average atom-molecule distance much larger than the orbit of a Rydberg electron. In this regime the interaction between molecules and Rydberg atoms is dominated by the van der Waals potential $V(r)=C_6/r^6$, where the $C_6$ is a state-dependent van der Waals coefficient. We recently computed  $C_6$ coefficients for a large set of alkali-metal Rydberg levels and selected alkali-dimers in their rovibrational ground state \cite{Olaya}.  For attractive atom-molecule potentials in an atomic Rydberg channel ($C_6<0$), we expect the laser excitation lineshape of the Rydberg level to be sensitive to the formation of long-range atom-molecule {\it trimer} bound states near threshold via two-photon photoassociation \cite{Munchow2011,Dutta2017}.

We compute the photoassociation (PA) spectrum and rates for a dilute ultracold mixture of ground state KRb molecules ($^1\Sigma^+,v=0, J=0$) in a reservoir gas of $^{85}$Rb atoms asymptotically prepared in the Rydberg channels $n^2S_{1/2}$ and $n^2D_{3/2}$, as a function of the principal quantum number $n$.  In what follows, we briefly describe the methodology used to compute the PA rates and lineshapes for selected free-to-bound transitions in Sec. \ref{sec:PA rates}, then discuss the results obtained at $T=1\,\mu{\rm K}$ in Sec. \ref{sec:results}. We emphasize the $n$-dependence of the PA process. Comparisons are made with other PA rates available in the literature for low-$n$ atom-atom and atom-molecule bound states formed near threshold. We conclude in Sec. \ref{sec:conclusion} with a discussion of possible scenarios for testing our predictions using currently available technology.

%We consider a two-photon Rydberg excitation of an alkali-metal atom in an ultracold bath of heteronuclear alkali-metal dimers as displayed in Fig. \ref{Fig_stellar}. In this way, we get rid of the characteristic gas heating associated with a single UV photon excitation. 

\section{Theoretical Framework}
\label{sec:PA rates}

As discussed above, the lineshape of the Rydberg excitation is directly related with the formation of excited trimer bound states supported by the underlying atom-molecule potential (see Fig. \ref{Fig_stellar}). The general theory of atom-atom photoassociation is reviewed in Ref. \cite{Jones}. The formalism is also applicable for atom-molecule systems in the van der Waals regime \cite{Blasing}. Here we focus on the PA rate that describes the laser-assisted binding of a $^{85}$Rb ground state atom ($5^2S_{1/2}$) and a KRb molecule in the ground rovibrational state ($^1\Sigma^+,v=0, J=0$) into a long-range atom-molecule trimer state that correlates asymptotically with the atomic Rydberg state $n^2L_j$, where $L$ is the atomic orbital angular momentum and $j$ the total electronic angular momentum. Spin-orbit coupling in the atomic Rydberg manifold is taken into account~\cite{Olaya}. The laser frequencies in the two-color excitation scheme are chosen such that the formation of long-range atom-molecule trimers  is most efficient at interparticle distances that far beyond the Le Roy radius of the asymptotic Rydberg state.

\subsection{Two-photon photoassociation rate}

The initial state of the PA process is a scattering wavefunction $|\Psi_l(E_{\text{kin}}) \rangle$ described by a collision energy $E_{\text{kin}}$ and partial wave $l$ in the ground atom-molecule channel $|5^2S_{1/2}\rangle |X^1\Sigma^{+} \rangle$. The  final state of the process $|\Psi_{\nu}\rangle$ corresponds to the $\nu$-th bound state of the trimer atom-molecule potential that correlates with the $|n^2L_{j}\rangle |X^1\Sigma^{+} \rangle$ asymptote. In this work we compute PA rates for $20\leq n \leq 80$ Rydberg levels in the $^2S_{1/2}$ and $^2D_{3/2}$ atomic channels, but the formalism can be equally applied to other atomic and molecular states, provided the $C_6$ coefficients in the initial and final states are known. From resonance scattering theory~\cite{Jones, Bohn}, the PA rate constant can be written as
\begin{equation}
\label{eq1}
K_{PA} (E_{\text{kin}}) =  v_{\text{rel}} \frac{\pi}{k^2} \sum_{l=0}^\infty (2l+1) |S_{\nu}(E_{\text{kin}},l)|^2,
\end{equation}

\noindent
where $v_{\text{rel}}= \sqrt{8 k_B T / \pi \mu}$ is the relative velocity of the particles in the entrance channel, $\mu$ is the reduced mass of the Rydberg atom-molecule system, and $k = \sqrt{2\mu k_B T / \hbar^2}$ is the channel wavevector. $k_B$ is the Boltzmann's constant and $T$ is the temperature. The rate constant is determined by the element of $S$-matrix that connects the scattering state and the bound state, given by~\cite{Jones}
\begin{equation}
\label{eq2}
|S_{\nu}(E_{\text{kin}},l)|^2 = \frac{ \gamma\, \Gamma_{\nu}(E_{\text{kin}},l) }{ [ E_{\text{kin}}/\hbar + \delta_\nu ]^2 + [ { \Gamma_{T}}/{2} ]^2}, 
\end{equation}
where $\delta_\nu=\omega_1+\omega_2-(E_a-E_\nu)$ is the two-photon detuning from the final state $|\Psi_{\nu}\rangle$, where $E_{a}$ is the energy of the Rb$^*$ ($n^2L_{j}$) + KRb ($X^1\Sigma^{+}, v=0, J=0$) asymptote, and $E_\nu>0$ is the binding energy of the $\nu$-th near-threshold bound state. For weak laser dressing, we can neglect possible light-shifts of these resonance frequencies. The two-photon excitation scheme is further specified in Fig. \ref{Fig_stellar} (right hand side) in terms of the individual Rabi frequencies $\Omega_1$ and $\Omega_2$ of the driving lasers at $\omega_1$ and $\omega_2$, respectively. The derivation of the effective two-level approximation used in Eq. (\ref{eq2}) is given in Appendix \ref{app:effective model}. 

The width of the scattering resonance in Eq.(\ref{eq2}) depends on the overall decay rate of the atom-molecule trimer state, given by $\Gamma_{\text{T}} = \gamma + \Gamma_{\nu}(E_{\text{kin}},l)$, where $\gamma$ is the natural linewidth of the Rydberg state and  $\Gamma_{\nu}(E_{\text{kin}},l) = 2\pi |V_{\nu} (E_{\text{kin}},l)|^2$ is the stimulated absorption rate associated with the the PA process. The latter is determined by
\begin{equation}\label{eq:overlap}
V_{\nu} (E_{\text{kin}},l) = \frac{\Omega_{\rm eff}}{2} \langle \Psi_{\nu} | \Psi_l(E_{\text{kin}}) \rangle,
\end{equation}
%, 
where $\langle \Psi_{\nu} | \Psi_l(E_{\text{kin}}) \rangle$ is the Franck-Condon Factor (FCF) between the initial scattering wavefunction and the final bound state wavefunction. For loosely bound vibrational states, it is a good approximation to neglect the radial dependence of the electric transition dipole moment, leading to an effective two-photon Rabi frequency of the form $\Omega_{\rm eff} = {\Omega_1 \Omega_2}/{2\Delta_1}$ (more details in Ref. ~\cite{Browaeys} and Appendix \ref{app:effective model}).

\subsection{Scattering and bound state wavefunctions}
The scattering wave function is obtained by numerically solving the Schr\"odinger equation for a Lennard-Jones potential describing the interaction of ground state KRb and ground state Rb. We solve the radial equation with the Numerov method for the Lennard-Jones parameters $C_6=8798$ au, $D_e=1684$ cm$^{-1}$, and $C_{12} \equiv C_6^2/4D_e$, taken from Ref. \cite{Mayle}. Scattering wave functions are obtained for a given collision energy $E_{\text{kin}}$ and partial wave $l$. For the ultracold temperatures of interest in this work, we only consider s-wave collisions ($l=0$). 

Bound state wavefunctions $| \Psi_{\nu} \rangle$ are calculated using a mapped Fourier grid Hamiltonian method~\cite{Kokoouline} given an atom-molecule trimer potential. This potential is in general unknown over the entire range of atom-molecule distances $R$. However, for distances beyond the Le Roy radius of the Rydberg atom $R_{\rm LR}$, it is reasonable to approximate the $s$-wave interaction potential by the van der Waals term  $V(r)=C_6/r^6$, provided that the $C_6$ coefficient for the atom-molecule pair is known. van der Waals coefficients for the interaction of KRb in the rovibrational ground state with a Rb atom in the $n^2L_j$ Rydberg level are taken from Ref.~\cite{Olaya}. We extend the van der Waals potential up to an arbitrary cutoff distance $R_c$ below the atomic Le Roy radius ($R_c<R_{\rm LR}$), where we then introduce a semi-infinite barrier.

\begin{figure}[t]
\centering
\includegraphics[width=0.48\textwidth]{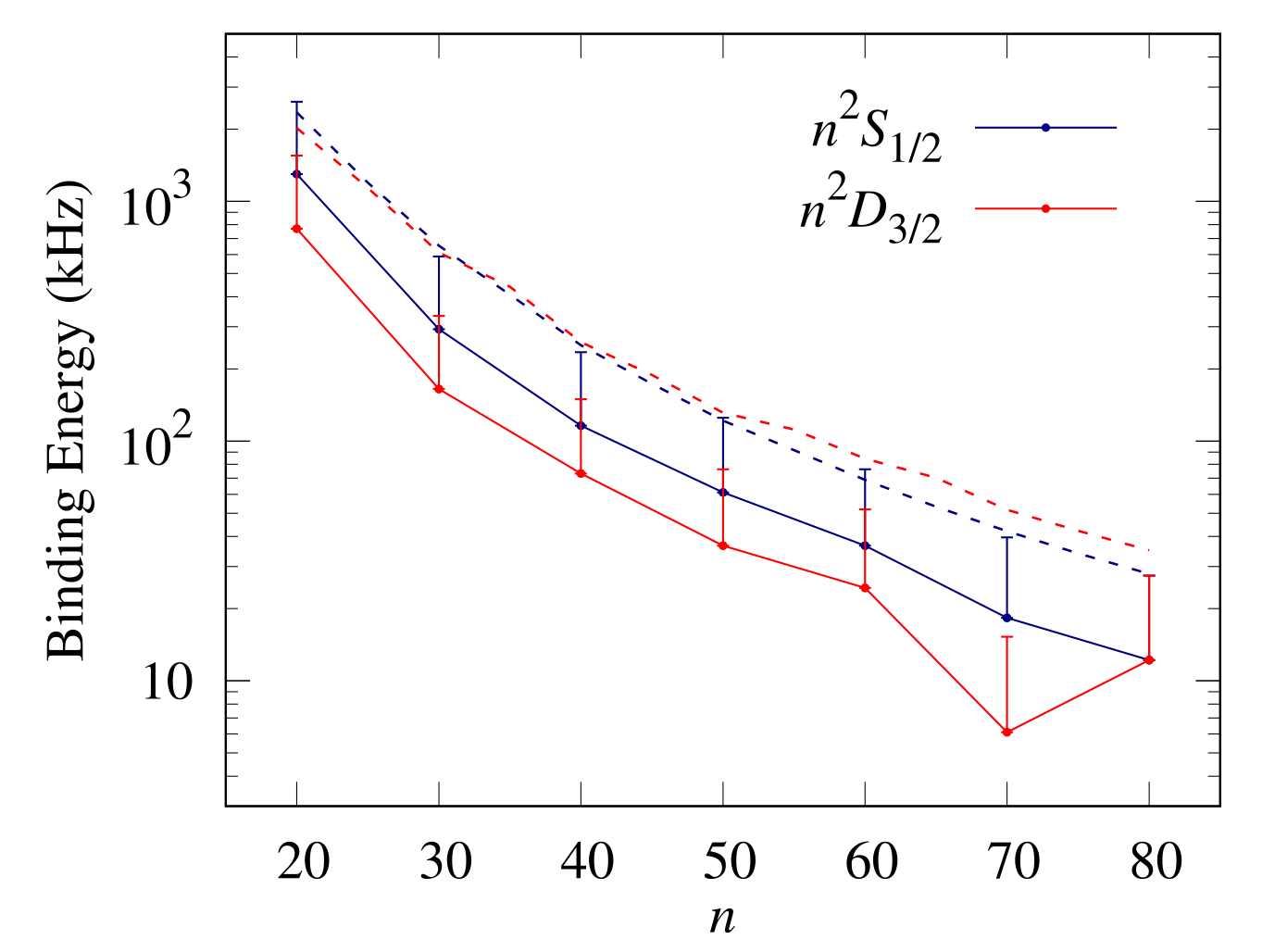}
\caption{{\bf Threshold trimer states of KRb-Rb$^*$}. Binding energy of the first bound state below threshold $E_{-1}$ for the atom-molecule trimer in the $n^2S_{1/2}$ and $n^2D_{3/2}$ asymptotes (solid lines), as a function of the atomic principal quantum number $n$. The dashed lines correspond to the analytical bound $39.5 E_{\rm vdW}$~\cite{Gao2000}, where the $E_{\rm vdW}$ is the characteristic energy of the van der Waals potential. Numerical error bars are obtained by changing the position of the short-range barrier below the atomic Le Roy radius.}
\label{fig:ene}
\end{figure}

The position of the short-range barrier $R_c$ is principle arbitrary, and it is known that near-threshold bound states can be very sensitive to the choice of $R_c$ \cite{Li2007}.  However, as shown first in Ref. \cite{Gao2000},  the binding energy of the first vibrational level below threshold $({\nu=-1})$ cannot be larger than $E_{-1}\approx 39.5\, E_{\rm vdW}$ in the $s$-wave regime, where $E_{\rm vdW}\equiv \sqrt{2}\hbar^3/\mu^{3/2}C_6^{1/2}$ is the characteristic energy of the van der Waals potential. Since we know the atom-molecule $C_6$ coefficients for the Rydberg channels of interest, we test the independence of our results to the choice of $R_c$ by comparing the numerical binding energy of the last bound state $\nu=-1$ with the analytical limit ($39.5\, E_{\rm vdW}$). The results are shown in Fig.  \ref{fig:ene}, for a broad range of principal quantum numbers  in the $n^2S_{1/2}$ and $n^2D_{3/2}$   channels. For every point we obtain an error bar that represents different choices of $R_c$. We confirm that the numerical binding energies $E_{-1}$ do not exceed the theoretical limit (dashed lines) and decrease with increasing $n$, following the scaling $C_6\sim n^7$ \cite{Olaya}. The numerical resolution of our bound state energies is about $3.1\,{\rm kHz}$, mostly constrained by the implementation of the Fourier grid~\cite{Kokoouline}. %However the last bound state energies are still sensible to the $C_6$ coefficient behaviour.

\section{Results}
\label{sec:results}

\begin{figure}[t]
\centering
\includegraphics[width=0.45\textwidth]{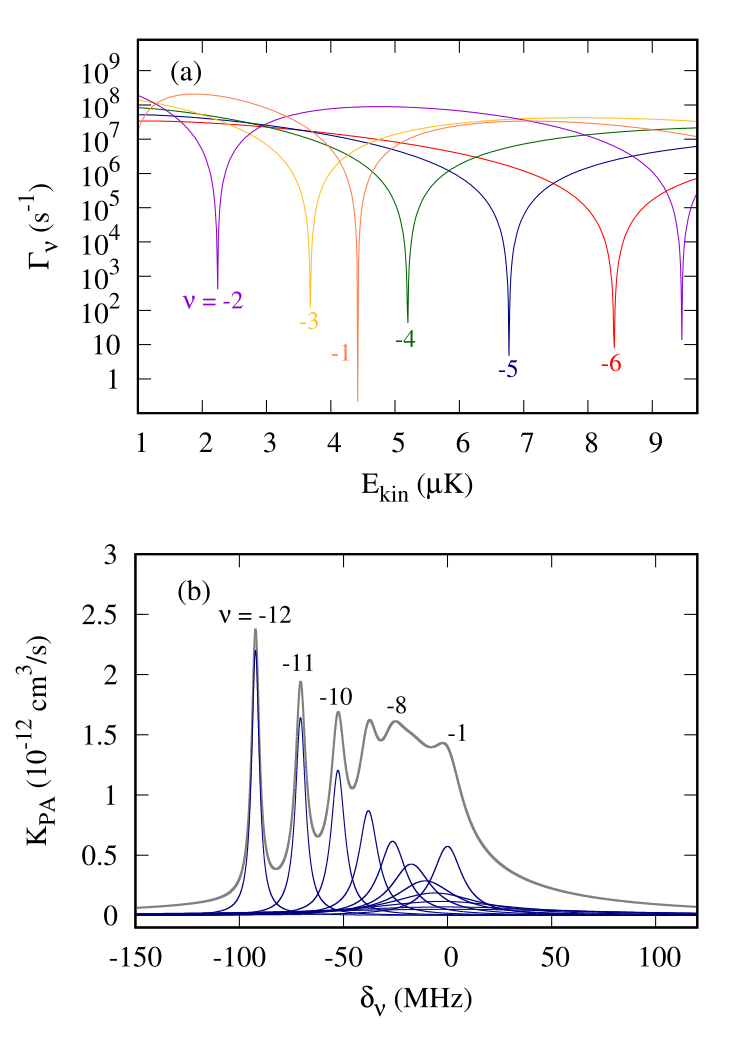}
\caption{{\bf PA spectrum for KRb-Rb$^*$ trimers}. (a) Stimulated absorption rate $\Gamma_{\nu}$ as a function of collision energy $E_{\text{kin}}$ for the $50\,^2D_{3/2}$ atomic channel. (b) Total rate constant $K_{PA}$ as a function of the effective two-photon detuning $\delta$ at $T=1\mu$K. The natural linewidth of of the Rydberg level is $\gamma = 1/126.53\, (\mu{\rm s})^{-1}$ \cite{Beterov}, and the effective Rabi frequency is $\Omega_{\rm eff}=500$kHz.}
\label{fig:2}
\end{figure}

\subsection{PA spectrum for the $50^2D_{3/2}$ Rydberg channel}

In Figure \ref{fig:2}a we show the predicted stimulated trimer formation rates $\Gamma_\nu$ as a function of collision energy, for the first bound states below threshold $\nu=-1,-2,\ldots, -6$, in the $50\,^2D_{3/2}$ atomic channel. This particular Rydberg level is chosen as a representative example. Since the stimulated rates depend on the Franck-Condon overlap $\langle \Psi_{\nu} | \Psi_l(E_{\text{kin}}) \rangle$, the dips in the curves reflect the energy-dependent mismatch of amplitudes between the scattering and the bound state wavefunctions. The exact position of the dips  in the energy axis depends on the location of the outer turning point of the vibrational bound states. Away from the mismatch, the predicted stimulated rates are on the order of $10^7$--$10^8$ s$^{-1}$ for Rydberg dressing parameters $\Delta_1 = 80$ MHz,  $\Omega_1 = 10$ MHz and $\Omega_2 = 8$ MHz, which correspond to the laser intensities $I_1= 10^{-4}$ W/cm$^2$ and $I_2 = 1.5$ W/cm$^2$ at frequencies $\omega_1$ and $\omega_2$, respectively. 

The predicted values of $\Gamma_{\nu}$ can be compared with PA rates measured for the formation of alkali-metal dimers from weakly excited (low $n$) alkali metal atoms. Experiments at  $100\, \mu$K  have measured stimulated rates of $ 1.5\times 10^8$ s$^{-1}$ for RbCs$^{*}$ \cite{Kerman} and $3.5\times 10^7$ s$^{-1}$ for  LiRb$^{*}$ \cite{Dutta}, using PA laser intensities of the order of $10^2$--$10^3$ W/cm$^{-2}$. These examples involve  species that interact also via van der Waals forces, but with values of $C_6$ that are several orders of magnitude smaller than atomic Rydberg channels. In other words, the PA resonance in low-$n$ systems occurs at relative short distances of a few hundreds of Bohr radii \cite{Jones}, which makes the process more sensitive to the details of the initial scattering wavefunction. %In our Rydberg-dimer system, the two-photon PA resonance can occur at several thousdands of Bohr radii, depending on the principal quantum number $n$, which is found to facilitate the free-to-bound wavefunction overlap a low temperatures.

\begin{figure}[t]
\centering
\includegraphics[width=0.45\textwidth]{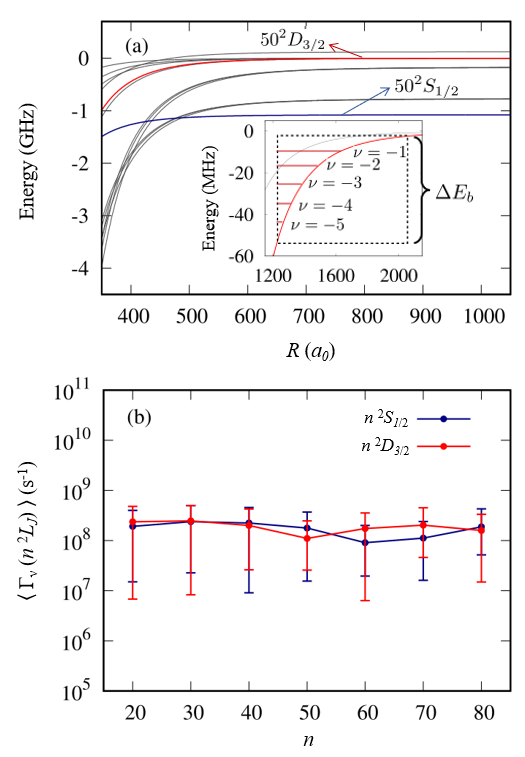}
\caption{{\bf PA rate dependence on the Rb$^*$ level}. (a) van der Waals potential for a KRb molecule ($X^1\Sigma^{+}, v=0, N=0$) interacting with an excited Rb$^*$ Rydberg atom. The potentials for the $50^2D_{3/2}$ and $50^2S_{1/2}$ channels are highlighted. The inset shows the highest bound trimer states within a range $\Delta E_b$ from threshold for $50^2D_{3/2}$. (b) Stimulated absorption rate $\langle \Gamma_{PA} \rangle$ as a function of the atomic principal quantum number $n$, averaged over bound states up to $ \Delta E_b = 100$MHz below threshold. The effective Rabi frequency is $\Omega_{\rm eff}=500$kHz and the collision energy is $E_{\text{kin}} = 1\mu$K.}
\label{fig:3}
\end{figure}

In Figure \ref{fig:2}b, we plot the PA rate constant $K_{\rm PA}$ for the formation of Rydberg-molecule trimers as a function of the two-photon detuning $\delta$, taking into account bound states that have a binding energy no greater than $100$ MHz. The collision energy is set to $E_{\rm kin}=1.0 \,\mu{\rm K}$ ($\approx 21$kHz). At such a low energies it is possible to detect the bound state resonances in the long-range tail of the Rydberg-molecule potential. The energy of the least bound state is $E_{-1}=37{\rm kHz}$, and the deepest bound state within the detunings of interest has a binding energy $E_{-12} = 94.6\,{\rm MHz}$, relative to the asymptotic energy of the $50\,^2D_{3/2}$ channel.

The predicted PA rate coefficient for the atomic Rydberg channel $50\,^2D_{3/2}$ is $K_{\rm PA}= 2.3 \times 10^{-12}$ cm$^{3}/$s for $\nu=-12$, which is smaller than the theoretical unitarity limit at $1 \mu$K ($3.07 \times 10^{-10}$ cm$^{3}/$s) obtained by setting $S_\nu(E,l)\rightarrow 1$ in Eq. (\ref{eq1}). This relatively large difference between the computed PA rate and the unitarity limit rate persists for other atomic channels, and is due to the small value of the natural linewidth of Rydberg levels $\gamma=1/\tau$ ($\tau=126.53 \mu$s \cite{Beterov}). The latter is at least  four orders of magnitude smaller than the predicted stimulated PA rates $\Gamma_\nu$ for the range of principal quantum numbers studied ($n=20-80$), thus suppressing the magnitude of the $S$-matrix element despite the potentially efficient transfer between the scattering state and the trimer bound state.

\subsection{PA rates for the $n^2S_{1/2}$ and $n^2D_{3/2}$ channels \\with $20\leq n\leq 80$}

In order to gain insight on the dependence of the PA transition rate $\Gamma_\nu$ with the principal quantum number $n$ of the atomic asymptote, we compute an average PA rate $\langle \Gamma_{\nu} \rangle$ over the set of bound states that lie within a predefined energy interval $\Delta E_b$ below a target asymptotic level $n^2L_j$. The results are shown in Fig.  \ref{fig:3}a for the $50^2S_{1/2}$ and $50^2D_{3/2}$ channels. Figure \ref{fig:3}a shows the corresponding potential energy curves near the outer turning points of the first bound states below threshold, where the bound state wavefunctions have the largest amplitudes. For the $50^2D_{3/2}$ channel shown in the inset, the outer turning points occur at about $1500 \,a_0$, which is an order of magnitude larger than in conventional PA for low-$n$ atomic scattering \cite{Dutta}.  

In Fig. \ref{fig:3}b, we plot the average  stimulated absorption rate $\langle \Gamma_{PA} \rangle$ as a function of the atomic principal quantum number $n$. The average is computed over all the bound states of the trimer with binding energy no greater than $\Delta E_b = 100$~MHz relative to the dissociation threshold in the $n^2S_{j}$  and $n^2D_{3/2}$ channels. In the average we also include a range of short-range barrier positions in the model Rydberg trimer potential. Figure \ref{fig:3}b shows that the average rate $\langle \Gamma_{PA} \rangle$  is largely insensitive to $n$. This is again due to the large $C_6$ coefficients of the trimer potential, which pushes the trimer bound state wavefunction to atom-molecule distances where the initial scattering wavefunction has a slowly-varying amplitude.

\section{Conclusion} 
\label{sec:conclusion}

In order to place our results in the context of previous photoassociation schemes for the formation of long-range bound states with low-$n$ atomic channels, consider the photoassociation of  RbCs$^{*}$ molecules starting from Rb and Cs ground states at $100\,\mu$K~\cite{Dutta}. The PA rate constant for this dimer system has a theoretical value $K_{\rm PA}\sim 10^{-11}$ cm$^{3}/$s, which is higher by an order of magnitude than the measured rate~\cite{Dutta}. Similar discrepancies between theory and experiments are found for the heteronuclear alkali dimers NaCs and LiK \cite{Dutta}. In contrast, for LiRb dimers  at $ 1\,{\rm mK}$, closer agreement between the experimental PA rate ($ 1.3\times 10^{-10}$ cm$^{3}/$s) and the unitarity limit ($ 2.1\times 10^{-10}$ cm$^{3}/$s) has been demonstrated \cite{Dutta}. The rate constants predicted in our work are thus comparable with those measured for low-$n$ systems. 

In summary, we study the van der Waals interaction between $^{85}$Rb atoms and KRb molecules in a two-photon photoassociation scheme that promotes the atom into a high-$n$ Rydberg level $n^2L_j$, for a molecule that remains in its rovibrational ground state.  The Rydberg excitation process can efficiently create bound trimer states with atom-molecule distances on the order of $ 10^3\, a_0$. These bound states are  supported by the strongly attractive $C_6$ coefficients of the collision pair in the atomic Rydberg channel \cite{Olaya}. The average stimulated absorption rates into a KRb-Rb$^{*}$ trimer is as high as $10^8\,{\rm s}^{-1}$ at 1 $\mu{\rm K}$, for easily accessible two-photon laser excitation parameters \cite{Guttridge2018}. Given the favorable overlap between the scattering and Rydberg-level bound state wavefunctions at such large distances, the unitarity limit of the transfer from the scattering state to the trimer bound states is within reach using atom-molecule co-trapping techniques \cite{Moses2017}, limited only by the relatively small natural linewidth of Rydberg atoms. 

In comparison with traditional cold-molecule formation experiments that target high-density molecular ensembles \cite{Matsuda2020}, our Rydberg trimer predictions could be tested in a low-density regime where independent molecules are diluted in a background gas of alkali-metal atoms, for atomic densities below the critical value beyond which Rydberg blockade effects become important \cite{Balewski}. Our work thus sets the foundations for a deeper understanding of exotic properties of ultracold molecules in atomic Rydberg reservoirs.

\section*{Acknowledgements}

 F.H. and V.O. thank support by ANID--Fondecyt Regular 1181743 and ANID--Millennium Science Initiative Program ICN17\_012.

\appendix
\section{Effective two-level system}
\label{app:effective model}

The Hamiltonian that describes a three-level system as the one shown in Figure \ref{Fig_stellar} is given by ($\hbar=1$)

\begin{equation}
\begin{aligned}
H &= \omega_e | e \rangle\langle e | + \omega_r | r \rangle\langle r | + \frac{\Omega_1}{2} \left[ |g \rangle\langle e| e^{i\omega_1 t} + |e \rangle\langle g| e^{-i\omega_1 t}  \right]\\
& \ \ \ \ + \frac{\Omega_2}{2} \left[ |e \rangle\langle r| e^{i\omega_2 t} + |r \rangle\langle e| e^{-i\omega_2 t}  \right],
\end{aligned}
\end{equation}

where $\omega_e$ ($\omega_r$) is the energy of the $|e \rangle$ ($|r \rangle$) state and the energy of $|g\rangle$ is set to zero. $\Omega_a$ and $\omega_a$ are the Rabi frequency and energy of the laser $a$, with $a=1,2$.

Using an unitary rotation frame transformation $U(t) = | g\rangle\langle g| + e^{i\omega_1 t} |e \rangle \langle e | + e^{i(\omega_1 + \omega_2)t} |r \rangle \langle r |$, the interaction Hamiltonian can be written as \cite{DanielBook}

\begin{equation}
\begin{aligned}
H_{I} &= - \Delta_1 | e \rangle\langle e | + \delta | r \rangle\langle r | \\
& \ \ + \frac{\Omega_1}{2} \left[ |g \rangle\langle e| + |e \rangle\langle g| \right] + \frac{\Omega_2}{2} \left[ |e \rangle\langle r|  + |r \rangle\langle e|   \right],
\end{aligned}
\end{equation}

where $\delta=\omega_r - (\omega_1 + \omega_2)$ and $\Delta_1 = \omega_1 - \omega_e$. The time evolution of the system is determined by the Schr\"odinger equation $i \partial_t |\Psi \rangle = H_{I} | \Psi \rangle$, where $|\Psi (t) \rangle = c_g(t) |g\rangle + c_e(t) |e\rangle + c_r(t) |r \rangle$ is the system state. In the rotating frame, $|e \rangle$ has fast oscillations and this state instantaneously tends to a steady state compared with the slow motion of the rest of the system, therefore we assume that $\dot{c}_e = 0$. Thus, the equations of motion are given by

\begin{equation}\label{eq:A1}
i\dot{c}_{g}(t) = \frac{\Omega_1^2}{4\Delta_1} c_g(t) + \frac{\Omega_1\Omega_2}{4\Delta_1} c_r(t),
\end{equation}

\begin{equation}\label{eq:A2}
i\dot{c}_r(t) = \delta c_r(t) + \frac{\Omega_2^2}{4\Delta_1} c_r(t) + \frac{\Omega_1\Omega_2}{4\Delta_1} c_g(t).
\end{equation}

Eqs. (\ref{eq:A1}) and (\ref{eq:A2}) are the same differential equation obtained from a two-level system with and effective Rabi frequency $\Omega_{\rm eff}= \frac{\Omega_1\Omega_2}{2\Delta_1}$ and an effective detuning

\begin{equation}
\delta_{\rm eff} = \delta + \frac{\Omega_2^2}{4\Delta_1} - \frac{\Omega_1^2}{4\Delta_1}.
\end{equation}

\bibliography{trimer}{}

\bibliographystyle{apsrev}

\end{document}